# Magnetic properties of monoclinic lanthanide orthoborates, *Ln*$BO_3$, *Ln* = Gd, Tb, Dy, Ho, Er, Yb

P Mukherjee[1*], Y Wu[1], G I Lampronti[2], S E Dutton[1*]

[1] Cavendish Laboratory, University of Cambridge, JJ Thomson Avenue, Cambridge CB3 0HE, United Kingdom

[2] Department of Earth Sciences, Downing Site, University of Cambridge, Downing Street, Cambridge CB2 3EQ, United Kingdom

*E-mail: pm545@cam.ac.uk, sed33@cam.ac.uk

## Abstract:

The lanthanide orthoborates, *Ln*$BO_3$, *Ln* = Gd, Tb, Dy, Ho, Er, Yb crystallise in a monoclinic structure with the magnetic $Ln^{3+}$ forming an edge-sharing triangular lattice. The triangles are scalene, however all deviations from the ideal equilateral geometry are less than 1.5%. The bulk magnetic properties are studied using magnetic susceptibility, specific heat and isothermal magnetisation measurements. Heat capacity measurements show ordering features at $T \leq 2$ K for *Ln* = Gd, Tb, Dy, Er. No ordering is observed for $YbBO_3$ at $T \geq 0.4$ K and $HoBO_3$ is proposed to have a non-magnetic singlet state. Isothermal magnetisation measurements indicate isotropic $Gd^{3+}$ spins and strong single-ion anisotropy for the other $Ln^{3+}$. The change in magnetic entropy has been evaluated to determine the magnetocaloric effect in these materials. $GdBO_3$ and $DyBO_3$ are found to be competitive magnetocaloric materials in the liquid helium temperature regime.

# 1 Introduction

Materials having magnetic lattices with triangular or tetrahedral geometries are often frustrated due to the inability of all the pairwise interactions to be satisfied simultaneously. For a particular frustrated plaquette, the magnetic properties vary widely because factors including crystal electric field (CEF) effects and lattice distortions compete with the magnetic interactions to determine the magnetic ground state. Depending on the relative magnitudes of such interactions, exotic ground states may emerge. Realisation of such states in real materials open up the possibility of testing theoretical predictions and realisation of novel magnetic properties[1–3].

There have been many studies on three-dimensional (3D) frustrated lattices containing magnetic $Ln^{3+}$ ions; most notably the pyrochlores - $Ln_2B_2O_7$ ($B$ = Ti, Sn) and more recently $Ln_2Zr_2O_7$ but also other materials including gadolinium gallium garnet (GGG) and the $SrLn_2O_4$ family of materials[4–11]. Work on two-dimensional (2D) frustrated lattices containing magnetic $Ln^{3+}$ has been limited due to the lack of experimental realisations[12], however this field is gaining momentum. Recently the isostructural series $Ln_3X_2Sb_3O_{14}$ ($X$ = Mg, Zn) have been reported which contain structurally perfect 2D kagome planes of magnetic $Ln^{3+}$[13–15]. Several exotic ground states have been already reported including umbrella-like all-in all-out long range ordering for $Nd_3Mg_2Sb_3O_{14}$, dipolar interaction mediated long-range ordering in a 120$^o$ structure for $Gd_3X_2Sb_3O_{14}$ ($X$ = Mg, Zn), emergent charge order in $Dy_3Mg_2Sb_3O_{14}$ and a possible Kosterlitz-Thouless (KT) vortex unbinding transition in $Er_3Mg_2Sb_3O_{14}$ [16–18]. Another recent discovery is the Quantum Spin Liquid (QSL) candidate $YbMgGaO_4$ where the magnetic $Yb^{3+}$ with effective spin $S$ = ½ form a triangular lattice. Bulk magnetic measurements in this material shows no evidence of ordering down to 60 mK while neutron scattering experiments have revealed a continuum of magnetic excitations, consistent with a QSL state[18,19]. The bulk magnetic properties of the $KBaLn(BO_3)_2$ series, which crystallise in a structure containing edge-sharing triangular lattices of $Ln^{3+}$, have also been recently reported [20]. Discovery of other 2D frustrated lattices with magnetic rare earth ions opens up the possibility of exploring further aspects of 2D geometrically frustrated systems.

Lanthanide orthoborates, $LnBO_3$, have been widely studied for their optical properties because they have high ultraviolet transparency and high optical damage thresholds, making them suitable for applications as phosphors in vacuum discharge lamps and screens[21–23]. However, except for early studies on magnetic susceptibility[24,25], their magnetic properties have not been explored. . The synthesis and crystal structure of the $LnBO_3$ was first reported by Levin *et. al*[26]. It was proposed that the lanthanide orthoborates $LnBO_3$ crystallise in the same three structures as $CaCO_3$ depending on the $Ln^{3+}$ ion - aragonite for the larger $Ln^{3+}$ (La - Nd), vaterite for the smaller $Ln^{3+}$ (Eu - Yb) and calcite for the smallest $Ln^{3+}$ ion, Lu. $SmBO_3$ was reported to crystallise in the vaterite phase between 1100 and 1300 $^o$ C and in a different triclinic structure at other temperatures. Since then, there has been much debate about the crystal structure of the so-called $\pi$-$LnBO_3$ with $Ln$ = Eu – Yb, with later studies proposing the existence of both a hexagonal or 'pseudo-vaterite' [22,27] and monoclinic structure[23,28,29]. However, in both the proposed structures, the magnetic $Ln^{3+}$ link to form edge-sharing triangles and thus, the $\pi$-$LnBO_3$ may be an example of a new series of geometrically frustrated magnetic materials containing $Ln^{3+}$.

In this paper we report the synthesis, characterisation and bulk magnetic properties on polycrystalline samples of *Ln*$BO_3$, *Ln* = Gd, Tb, Dy, Ho, Er and Yb. The materials were prepared by solid state synthesis and the crystal structure was analysed using powder X-Ray diffraction (PXRD). The bulk magnetic properties have been studied using magnetic susceptibility, heat capacity and isothermal magnetisation measurements. To the best of our knowledge, this is the first comprehensive report on the magnetic properties. The *Ln*$BO_3$ exhibit different magnetic ordering features, the magnetic behaviour is highly dependent on the $Ln^{3+}$ under consideration. Evaluation of the magnetocaloric effect shows that $GdBO_3$ and $DyBO_3$ are competitive materials for solid state magnetic refrigeration in the liquid helium temperature regime, $T \geq 2$ K.

## 2 Experimental Section

Samples of *Ln*$BO_3$, *Ln* = Gd, Tb, Dy, Ho, Er and Yb, were prepared using a solid-state synthesis method. $Gd_2O_3$ was pre-dried at 800 $^oC$ overnight prior to being weighed out to ensure the correct stoichiometry. Samples were prepared by mixing stoichiometric amounts of $Ln_2O_3$ (*Ln* = Gd, Dy, Ho, Er, Yb) or $Tb_4O_7$ and $H_3BO_3$ (5% excess to compensate for the loss of B due to volatilisation during heating). A pre-reaction was carried out at 350 $^oC$ for 2 hours to decompose the $H_3BO_3$ to $B_2O_3$. After regrinding, samples were heated to 1000 $^oC$ for either 24 or 48 hours to obtain the final product.

The formation of a phase pure product was confirmed using room temperature (RT) powder X-Ray diffraction (PXRD). Initially short scans were collected over $5^o \leq 2\theta \leq 60^o$ ($\Delta\ 2\theta$ = 0.015$^o$) using a Panalytical Empyrean X-Ray diffractometer (Cu Kα radiation, λ = 1.541 Å). For more detailed structural analysis, longer scans at high resolution were collected using a Bruker D8 Advance diffractometer (Cu Kα radiation, λ = 1.541 Å, Ge monochromator and Sol-XE energy dispersive detector). Measurements were carried out for a day over an angular range $10^o \leq 2\theta \leq 120^o$ ($\Delta\ 2\theta = 0.01^o$). Rietveld refinement was carried out using the Fullprof suite of programs[30]. The backgrounds were fitted using linear interpolation and the peak shape was modelled using a pseudo-Voigt function.

Magnetic susceptibility measurements were performed on a Quantum Design Magnetic Properties Measurement System (MPMS) with a Superconducting Quantum Interference Device (SQUID) magnetometer. The zero-field cooled (ZFC) susceptibility $\chi(T)$ was measured in a field of 100 Oe in the temperature range 2-300 K. In a field of 100 Oe, the isothermal magnetisation $M(H)$ curve is linear at all $T$ and so $\chi(T)$ can be approximated by the linear relation: $\chi(T) \sim M/H$. Isothermal magnetisation, $M(H)$, measurements in the field range, $\mu_o H = 0 - 9$ T for selected temperatures were carried out using the ACMS (AC Measurement System) option on a Quantum Design Physical Properties Measurement System (PPMS)

Zero field heat capacity (HC) measurements were carried out using the He3 option on a Quantum Design PPMS in the temperature range 0.4 – 20 K. To improve thermal conductivity at low temperatures, samples were mixed with approximately equal amounts of silver powder (99.99%, Alfa Aesar). The contribution of the silver powder to the heat capacity was then deducted using values reported in the literature[31] to obtain the contribution to the heat capacity from the sample only. The lattice heat capacity was subtracted using a Debye model[32] to get the magnetic contribution, $C_{mag}(T)$.

# 3 Results and discussion

## 3.1 Crystal Structure

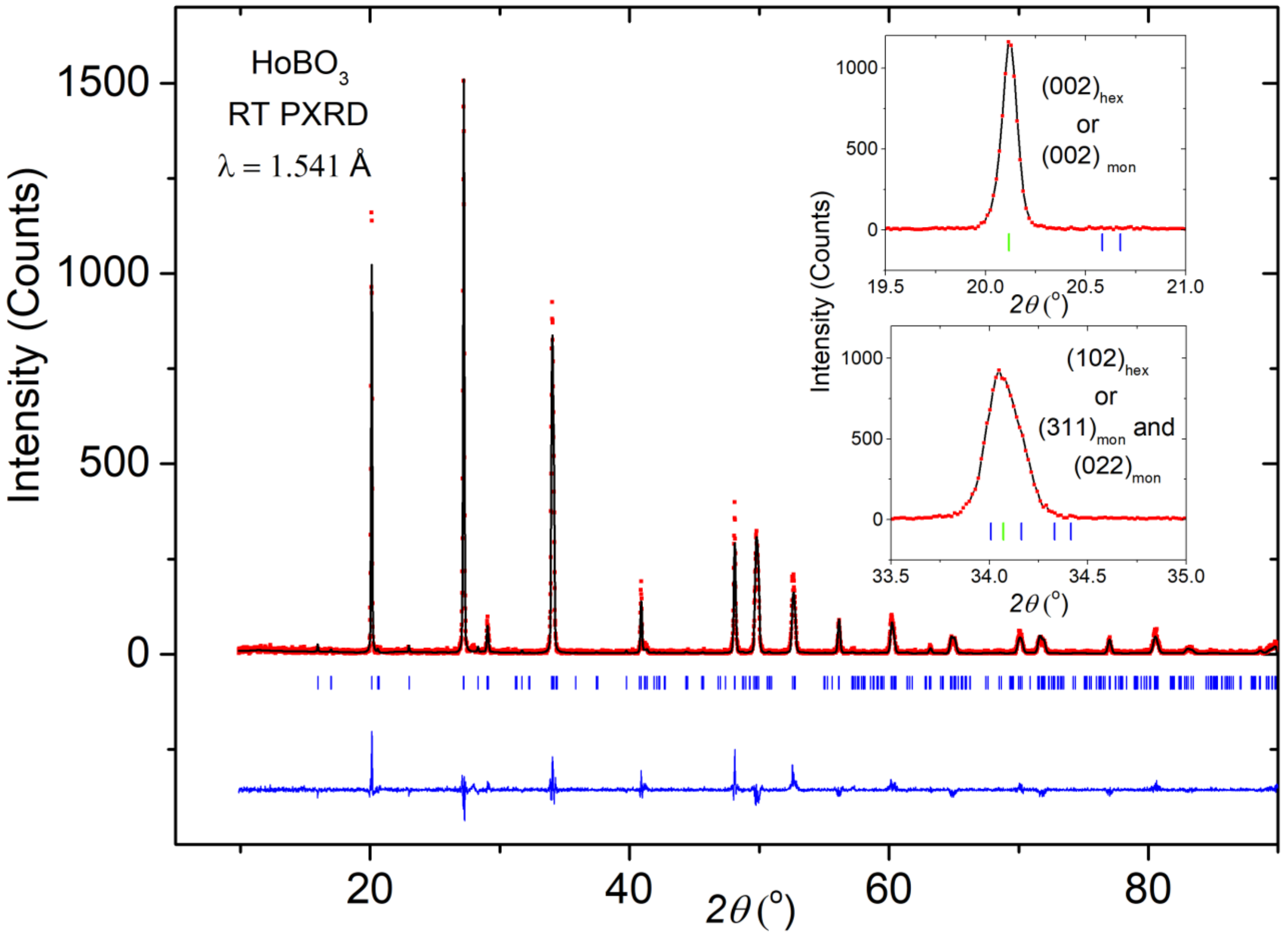

Figure 1 –PXRD pattern for $HoBO_3$: Experimental data (red dots), Modelled data (black line), Difference pattern (blue line), Bragg positions (blue ticks, reflections for hexagonal setting are highlighted in green); Inset: Peaks at 20.1 ° and 34.1°. Peak at ~ 20.1 ° corresponds to single reflection in both hexagonal $(002)_{hex}$ and monoclinic setting, $(002)_{mon}$. Peak at ~ 34.1 ° corresponds to single reflection in hexagonal setting, $(102)_{hex}$, but two reflections in monoclinic setting, $(311)_{mon}$ and $(022)_{mon}$

PXRD indicated formation of phase pure samples of $LnBO_3$, $Ln$ = Gd, Tb, Dy, Ho, Er and Yb. Attempts to synthesise $LnBO_3$ with larger $Ln^{3+}$ ions resulted in the formation of the orthorhombic or triclinic polymorphs as reported in the literature[26] and so will not be discussed here. The PXRD pattern for $HoBO_3$ is shown in Figure 1. We find that the intensities of the peaks for the π-$LnBO_3$ are not correctly modelled by the hexagonal phase and the monoclinic phase is most appropriate to describe the structure. This can also be confirmed by comparing the peak shape for selected peaks. Figure 1a inset for $HoBO_3$ shows two peaks at 20.1 ° and 34.1 °; the former corresponds to a single reflection in hexagonal setting, $(002)_{hex}$, and monoclinic setting, $(002)_{mon}$, while the latter corresponds to single reflection in hexagonal setting, $(102)_{hex}$, but two reflections in monoclinic setting, $(311)_{mon}$ and $(022)_{mon}$. The peak shape for the latter is consistent with a lower symmetry monoclinic structure. Thus the materials prepared using our synthetic route are definitely monoclinic for all *Ln*. In further discussions, we will use the term monoclinic $LnBO_3$ to refer to these compounds.

Details of the Rietveld analysis for the monoclinic $LnBO_3$ are given in Table 1. In our X-Ray analysis, the lattice parameters and the *Ln* positions were refined but the B and O positions were fixed to those reported in the literature for the monoclinic structure[23,28,29]. This is because PXRD is not sensitive to B and O in the presence of heavy *Ln*. The lattice parameters *a*, *b*, *c*, the in-plane area, $ac\sin\beta$, and the lattice volume, $abc\sin\beta$, for the monoclinic $LnBO_3$ (space group *C*2/*c*) all follow a linear relationship with ionic radii of the lanthanide ions[33], select plots shown in Figure 2. Our structural model has eclipsed triangular $Ln^{3+}$ layers separated by sheets of three membered-rings of corner sharing $BO_4^{5-}$ tetrahedra forming isolated $B_3O_9^{9-}$ units[28,29]; however it is not possible to be definitive about the arrangement of the borate units from the PXRD data alone. The proposed crystal structure is shown in Figure 3a.

Table 1 – Crystal structure parameters for monoclinic $LnBO_3$ - space group *C*2/*c*: *Ln*1 occupies the 4*c* site (0.25, 0.25, 0), *Ln*2 occupies the general 8*f* (*x*, *y*, *z*) site. The *Ln*2 positions were refined from the PXRD data. All the B and O positions were kept fixed as follows: B1 occupies the general 8*f* (*x*, *y*, *z*) site = (0.12011, 0.03790, 0.24691) while B2 occupies the 4*e* site (0, *y*, 0.25) = (0, 0.67520, 0.25). O1, O2, O3, O4 all occupy the general 8*f* (*x*, *y*, *z*) positions; O1 = (0.12550, 0.09200, 0.10199), O2 = (0.22293, 0.09316, 0.38870), O3 = (0.04837, 0.56643, 0.39233), O4 = (0.39142, 0.30823, 0.25174) while O5 occupies the 4*e* site (0, *y*, 0.25) = (0, 0.135, 0.25). The values of thermal parameters were kept fixed at $B_{iso}$ = 0.8 Å$^2$ for all atoms.

| ***Ln*** | | **Gd** | **Tb** | **Dy** | **Ho** | **Er** | **Yb** |
|---|---|---|---|---|---|---|---|
| ***a* (Å)** | | 11.4968(6) | 11.4299(4) | 11.3755(3) | 11.3357(4) | 11.2911(3) | 11.2006(2) |
| ***b* (Å)** | | 6.6402(3) | 6.6037(2) | 6.5757(3) | 6.5502(2) | 6.5236(2) | 6.47250(14) |
| ***c* (Å)** | | 9.6796(4) | 9.6408(2) | 9.6092(3) | 9.5776(2) | 9.5475(2) | 9.4901(2) |
| ***β* (Å)** | | 113.048(5) | 112.945(2) | 112.919(2) | 112.930(2) | 112.914(2) | 112.8116(2) |
| **Volume (Å$^3$)** | | 679.96(6) | 670.11(3) | 662.04(4) | 654.95(3) | 647.76(2) | 634.18(2) |
| **$\chi^2$** | | 1.12 | 1.04 | 1.39 | 1.55 | 1.54 | 1.41 |
| ***Ln*2: 8*f* (*x*,*y*,*z*)** | *x* | 0.0816(15) | 0.0811(8) | 0.0826(12) | 0.0819(6) | 0.0851(3) | 0.0839(4) |
| | *y* | 0.25282(19) | 0.25387(8) | 0.25351(10) | 0.25405(5) | 0.25280(5) | 0.25346(5) |
| | *z* | 0.49251(13) | 0.49894(11) | 0.49614(14) | 0.49743(8) | 0.50038(6) | 0.49993(6) |

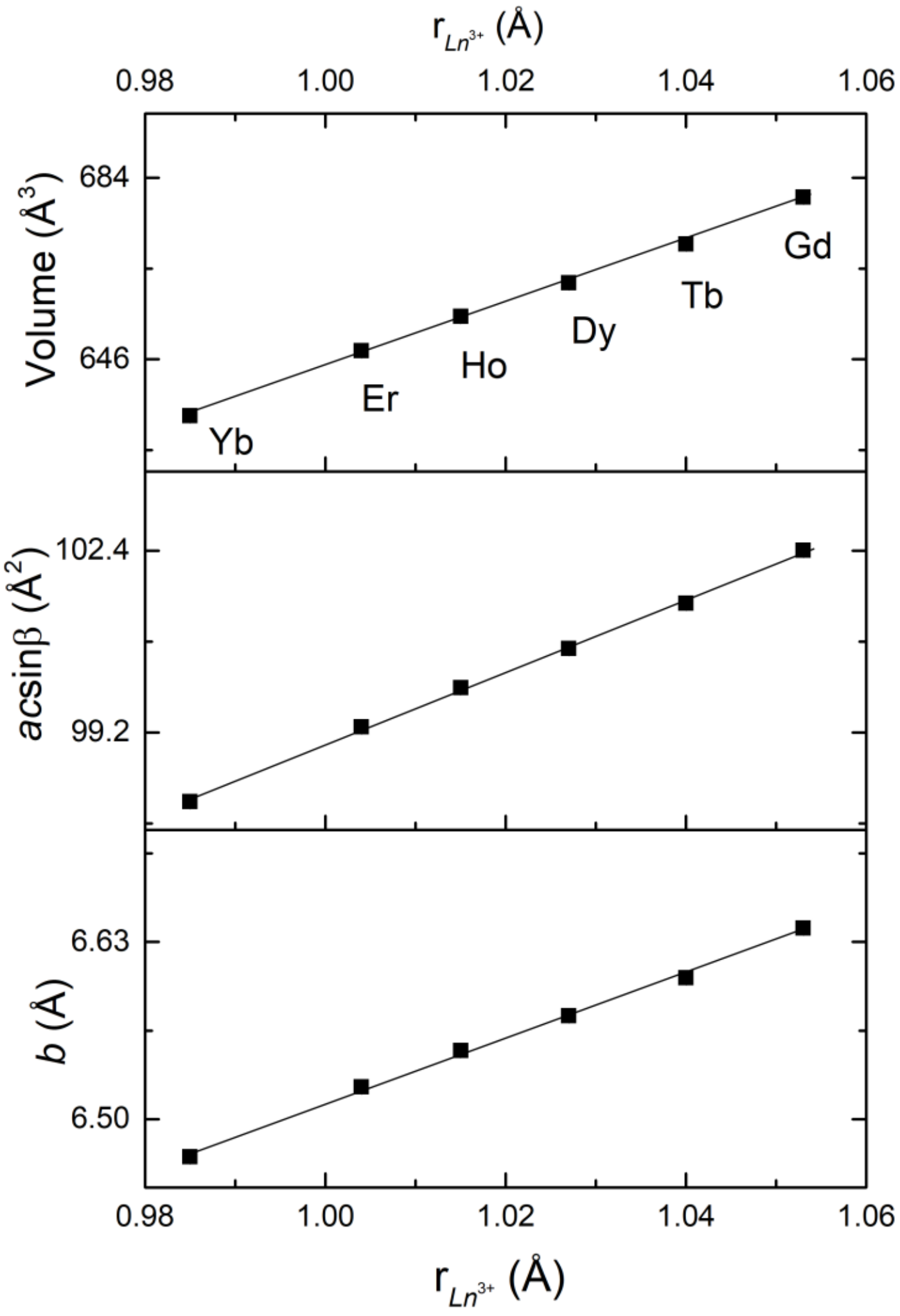

Figure 2 – Lattice parameters as a function of ionic radii for monoclinic *Ln*$BO_3$; from larger to smaller: *Ln* = Gd, Tb, Dy, Ho, Er, Yb. The plots show lattice parameter *b*, the planar area = $ac\sin\beta$ and cell volume = $abc\sin\beta$; lines are a guide to the eye

Determining whether the structure is hexagonal or monoclinic is significant for understanding the geometry of the frustrated magnetic lattice in these materials. If the structure were hexagonal, the magnetic $Ln^{3+}$ would have formed a perfectly flat lattice of edge-sharing equilateral triangles. However the monoclinic symmetry results in scalene triangles, with the triangular layers having a slight pucker which gives rise to two different interlayer *Ln*-*Ln* distances, *Ln*1-*Ln*2 and *Ln*2-*Ln*2 (Table S1). Three different sets of triangles comprising six different bond lengths (Table S1) are obtained, this is shown in Figure 3b. The distortion from the ideal two-dimensional triangular lattice in the monoclinic *Ln*$BO_3$ can be quantified by measuring the deviation in bond angles from ideal equilateral geometry; this is shown in Figure 4 for a single triangle. The deviations are identical for all three sets of triangles, Figure S1. Whilst the deviations from a perfect equilateral triangle are small, <1.5%, deviations of similar magnitude have been shown to have a dramatic effect on the magnetic properties of other geometrically frustrated triangular lattice systems[34–37].

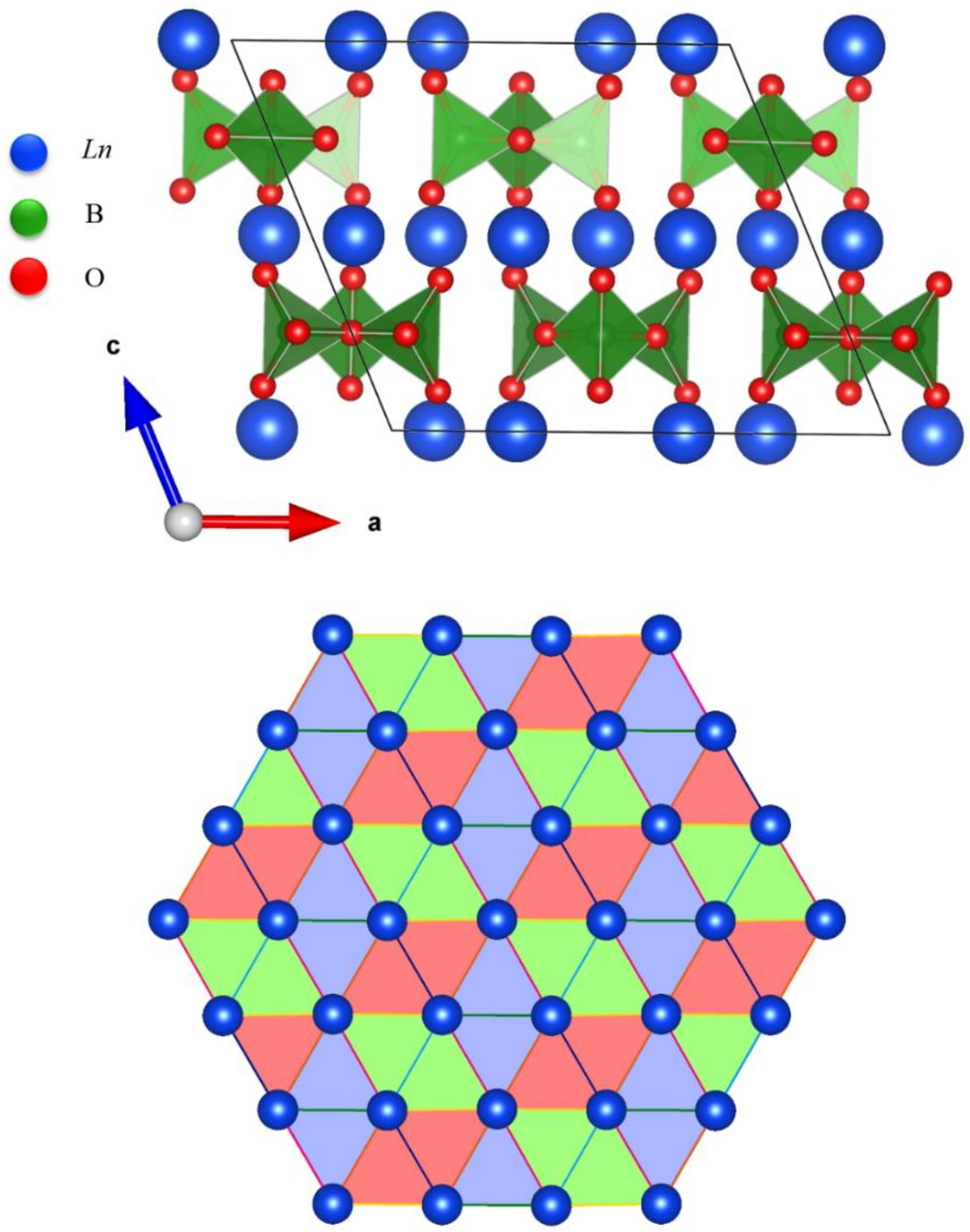

Figure 3 – a) Crystal structure of monoclinic *Ln*$BO_3$ with triangular sheets of $Ln^{3+}$ separated by layers of $B_3O_9^{9-}$ units b) Connectivity of magnetic $Ln^{3+}$ in monoclinic *Ln*$BO_3$ showing three different kinds of edge-sharing triangles. The different colours represent different bond lengths

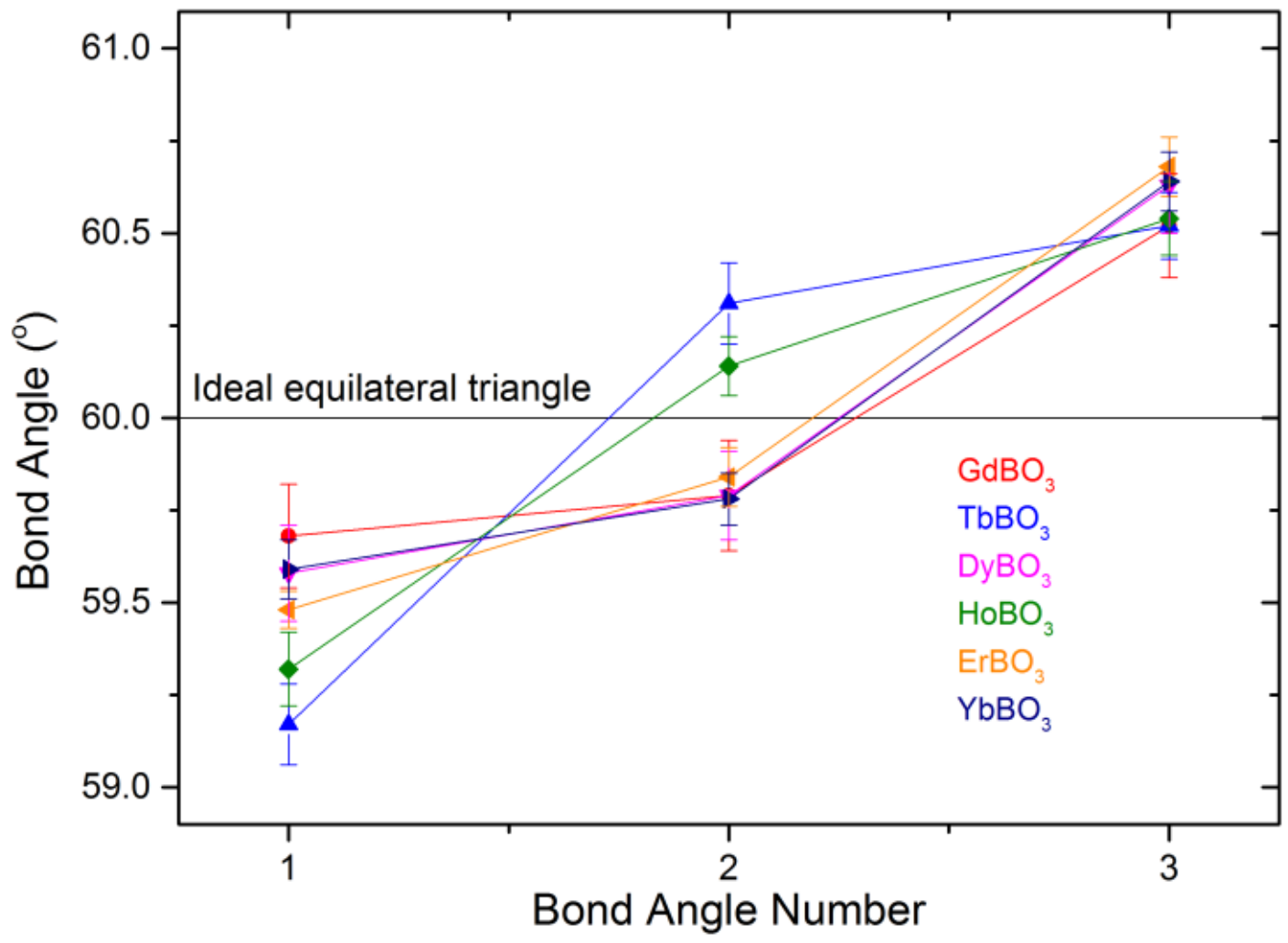

Figure 4 – Bond angles for a particular triangle in monoclinic *Ln*$BO_3$, *Ln* = Gd, Tb, Dy, Ho, Er, Yb. All the samples show deviation from ideal equilateral triangle; the % distortion is less than 1.5 %

## 3.2 Bulk magnetic measurements

### 3.2.1 Magnetic Susceptibility

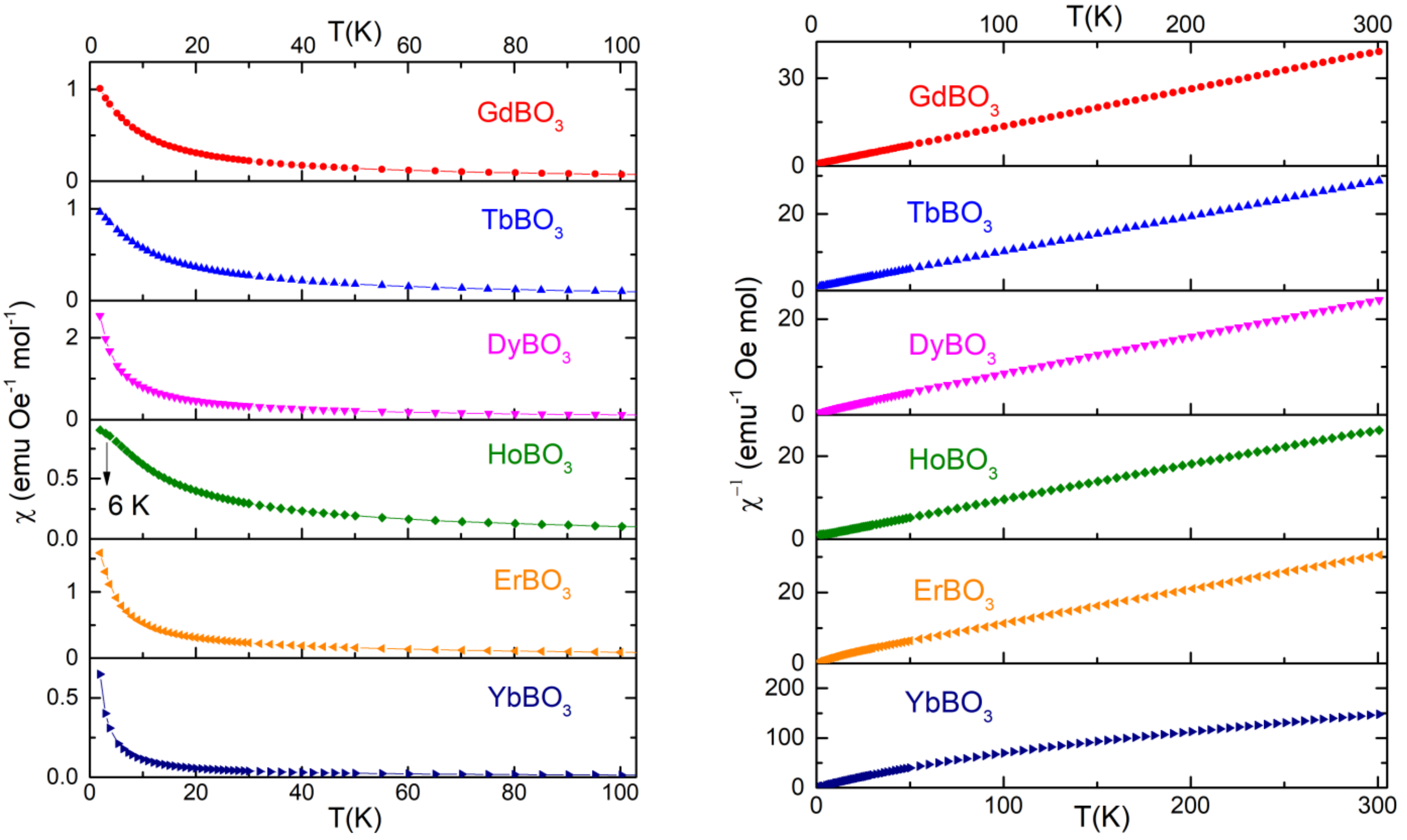

Figure 5 – a) ZFC $\chi(T)$ for monoclinic *Ln*BO3, *Ln* = Gd, Tb, Dy, Ho, Er, Yb b) Reciprocal of the molar susceptibility $\chi^{-1}(T)$ from 2 -300 K

The ZFC magnetic susceptibility, $\chi(T)$, for monoclinic *Ln*BO3 as a function of temperature (2-300 K) in a field of 100 Oe are shown in Figure 5. HoBO3 shows a feature at $T$ = 6 K; no ordering is observed down to 2 K for the other *Ln*BO3. The reciprocal susceptibility, $\chi^{-1}(T)$ is linear at temperatures $T$ >100K. Fits to the Curie-Weiss law $\chi^{-1} = \frac{T-\theta_{CW}}{C}$ were carried out in different temperature regimes from 100-300 K where $\theta_{CW}$ is the Curie-Weiss temperature and $C$ is the Curie constant. The average values were taken to calculate the experimental magnetic moment, $\mu_{eff} = \sqrt{\frac{3k_B C}{N_A \mu_B^2}}$ and $\theta_{CW}$. YbBO3 shows significant temperature-independent paramagnetism $\chi_0$, at higher temperatures due to CEF effects. Thus the Curie-Weiss fit for YbBO3 is carried out in the low temperature range, 2 – 30 K. The Curie-Weiss parameters are summarised in Table 2. The values of $\mu_{eff}$ are consistent with the theoretical values. The negative values of $\theta_{CW}$ indicate presence of antiferromagnetic correlations for all $Ln^{3+}$.

Table 2 – Parameters from bulk magnetisation measurements for monoclinic *Ln*$BO_3$ (*Ln* = Gd – Yb)

*Frustration index defined for transition at higher temperature

| **Compound** | **Theoretical $\mu_{eff}$ ($\mu_B$)** | **Experimental $\mu_{eff}$ ($\mu_B$)** | **$\theta_{CW}$ (K)** | **$T_0$(K)** | **$f$ = $\lvert\theta_{CW}/T_0\rvert$** | **Theoretical $M_{sat} = g_J J$ ($\mu_B$/f.u.)** | **$M_{max}$ at 2 K, 9T ($\mu_B$/f.u.)** |
|---|---|---|---|---|---|---|---|
| **$GdBO_3$** | 7.94 | 7.904 (4) | -5.4 (2) | 0.61, 1.72 | 3.1* | 7.0 | 6.6 |
| **$TbBO_3$** | 9.72 | 9.35 (2) | -11.0 (7) | 2.02 | 5.5 | 9.0 | 4.6 |
| **$DyBO_3$** | 10.65 | 10.19 (2) | -11.8 (6) | 0.56, 1.01 | 11.7* | 10.0 | 5.2 |
| **$HoBO_3$** | 10.61 | 9.73 (5) | -14 (2) | - | - | 10.0 | 4.8 |
| **$ErBO_3$** | 9.58 | 9.12 (3) | -18 (1) | 0.88 | 20.4 | 9.0 | 4.1 |
| **$YbBO_3$** | 4.54 | 3.058 (8) | -0.28 (9) | <0.4 | > 0.7 | 4.0 | 1.7 |

### 3.2.2 Isothermal magnetisation

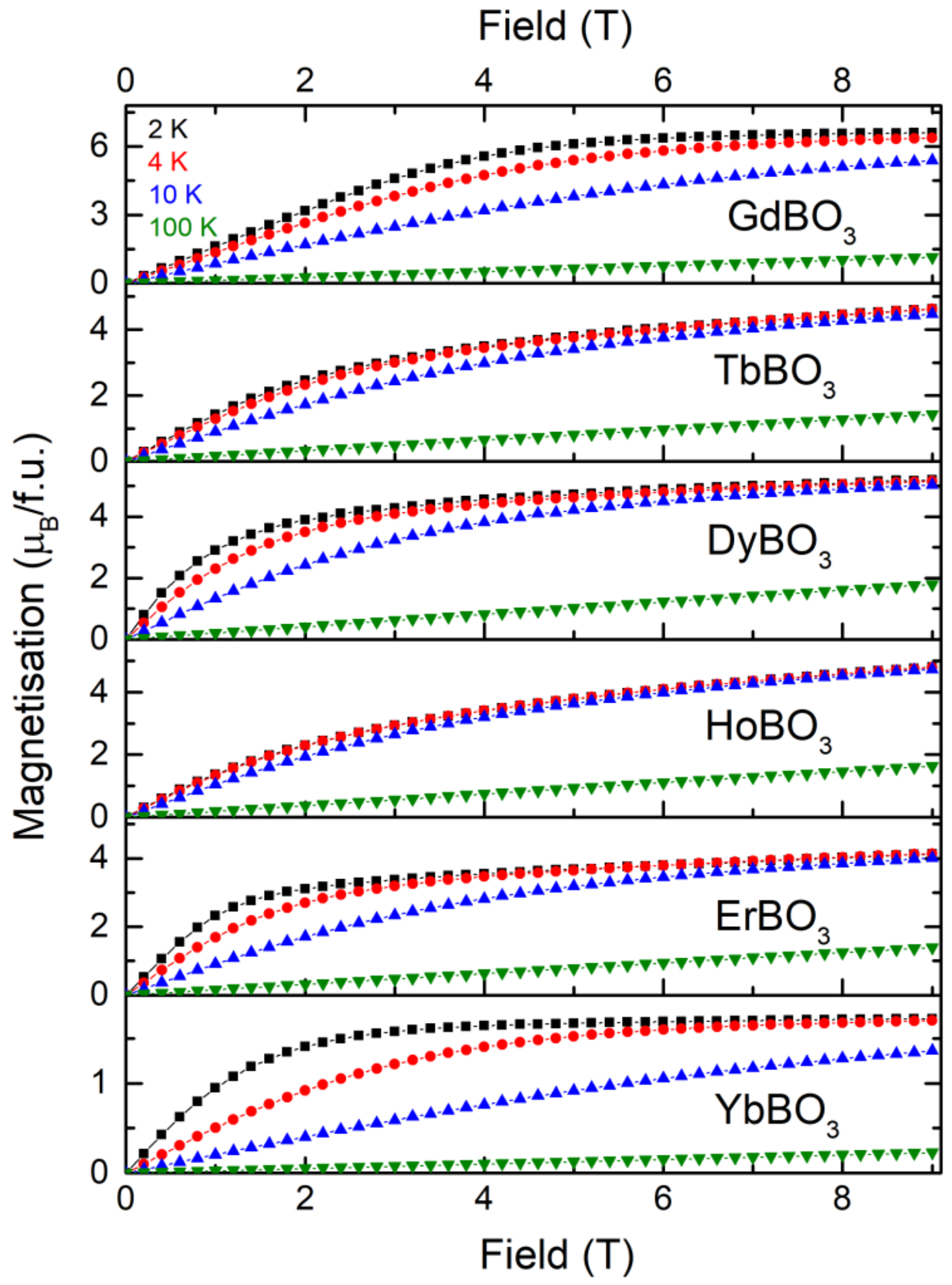

Figure 6 – Isothermal magnetisation $M(H)$ for monoclinic *Ln*$BO_3$, *Ln* = Gd, Tb, Dy, Ho, Er, Yb at selected temperatures in the magnetic field range 0 -9 T

The isothermal magnetisation curves for the monoclinic *Ln*$BO_3$ as a function of magnetic field at different temperatures are given in Figure 6. $GdBO_3$ saturates in a field of 9 T at 2 K; the maximum value = 6.6 $\mu_B$/f.u is consistent with the theoretical saturation magnetisation for Heisenberg spins, $M_{sat} = g_J J = 2\times7/2 = 7$ $\mu_B$/f.u. *Ln*$BO_3$, *Ln* = Tb, Dy, Ho, Er, do not show any signs of saturation in a field of 9 T. However, the values of maximum magnetisation at 2 K, 9 T, $M_{max}$ (Table 2), are consistent with previous reports for other 2D and 3D frustrated systems containing these magnetic ions[14,38]. We postulate that the $Ln^{3+}$ in *Ln*$BO_3$, *Ln* = Tb, Dy, Ho, Er, exhibit substantial single-ion anisotropy. $YbBO_3$ saturates at 2 K, 9 T; the maximum value, 1.7 $\mu_B$/f.u is close to $M_{sat}/2$, as reported for other frustrated systems containing edge sharing triangles of $Yb^{3+}$ where CEF effects lead to single-ion anisotropy[19,20]. Further experiments to determine the CEF levels are required to determine the nature of the single-ion anisotropy for *Ln* = Tb, Dy, Ho, Er, Yb.

### 3.2.3 Heat capacity

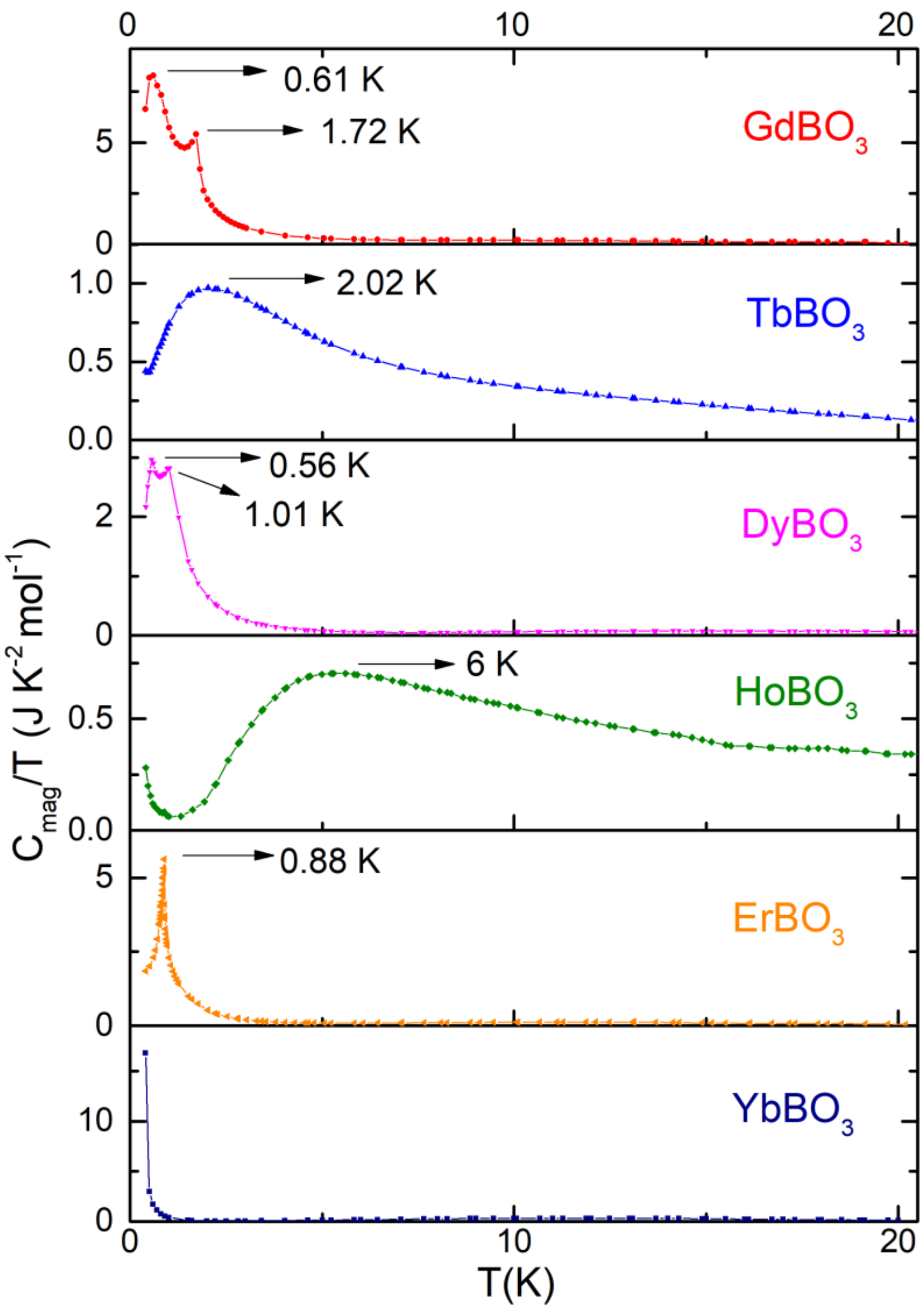

Figure 7 – Magnetic heat capacity $C_{mag}/T$ vs $T$ in zero field from 0.4 – 20 K for *Ln*$BO_3$, *Ln* = Gd, Tb, Dy, Ho, Er, Yb

We have carried out zero field heat capacity measurements to investigate the existence of magnetic ordering transitions for $T \geq 0.4$ K. Figure 7 shows the plot of $C_{mag}/T$ vs $T$ in zero field from 0.4 – 20 K for $Ln$BO$_3$, $Ln$ = Gd, Tb, Dy, Ho, Er, Yb, where $C_{mag}$ is the magnetic heat capacity.

$GdBO_3$ shows two sharp ordering transitions at 0.61 K and 1.72 K; as does $DyBO_3$ at 0.56 and 1.01 K. $ErBO_3$ shows a single $\lambda$ type transition at 0.88 K. No ordering for $YbBO_3$ is seen down to 0.4 K, however as $T$ approaches 0.4 K, a sharp increase in $C_{mag}$ can be seen, indicative of the onset of ordering at $T < 0.4$ K. A single broad feature is seen for $TbBO_3$ at 2.02 K, indicative of short range magnetic ordering. For $HoBO_3$, a very broad feature is seen at 6 K. This is likely due to van Vleck paramagnetism from the ground state being a non-magnetic singlet due to the non-Kramer's nature and low symmetry of $Ho^{3+}$, similar to the frustrated double perovskite $Ba_2HoSbO_6$ where the $Ho^{3+}$ ions lie on a fcc lattice[39]. Inelastic neutron spectroscopy and crystal electric field (CEF) calculations are needed to confirm this hypothesis. The upturn below 1.25 K is attributed to the nuclear Schottky anomaly for $Ho^{3+}$ which dominates the heat capacity at temperatures < 1 K as has been reported for other compounds[39–41]. The frustration index has been calculated according to the criterion proposed by Ramirez[42] for $Ln$BO$_3$, $Ln$ = Gd, Tb, Dy, Er. $HoBO_3$ has a non-magnetic ground state while a lower limit, $f > 0.7$, is obtained for $YbBO_3$.

The nature of magnetic ordering and the degree of frustration are different for $Ln$ = Gd, Tb, Dy, Ho. We postulate that the differences in magnetic interactions, crystal electric field effects (CEF) and lattice distortions result in different types of magnetic ordering for the various $Ln^{3+}$, as has been reported for other frustrated lanthanide oxide systems[4,17]. The observation of two magnetic ordering features, as in $GdBO_3$ and $DyBO_3$, has been observed in other frustrated Heisenberg systems like $SrGd_2O_4$[43], $Gd_2Ti_2O_7$[44] as well as for Ising systems including $Ca_3Co_2O_6$[45] and $CoNb_2O_6$[46] and $SrHo_2O_4$[10]. Usually in these systems, the transition at lower temperature is due to reorientation of spins, however further experiments are required to determine the origin of the magnetic ordering in $GdBO_3$ and $DyBO_3$. The sharp $\lambda$ type anomaly in $ErBO_3$ points to three-dimensional antiferromagnetic ordering, as has been reported for $SrEr_2O_4$[47] while the broad feature in $TbBO_3$ is reminiscent of short-range magnetic correlations as reported for members of the Sr$Ln_2$O$_4$ family[10,11,48]. Further neutron scattering experiments and theoretical modelling of the relevant interactions are needed to understand the fundamental magnetic behaviour of the monoclinic $Ln$BO$_3$.

#### 3.2.4 Magnetocaloric Effect (MCE)

Geometrically frustrated systems containing $Ln^{3+}$ can also find practical applications in solid state magnetic refrigeration in the liquid helium temperature regime because of the suppressed magnetic transition temperatures and the large amount of magnetic entropy that can be extracted[49,50]. Solid state magnetic cooling utilises adiabatic demagnetisation refrigeration, which is based on the principle of the magnetocaloric effect (MCE) in magnetic materials[51,52]. We examine the performance of the monoclinic $Ln$BO$_3$ as magnetocaloric materials (MCMs) for low temperature magnetic cooling, $T \geq 2$ K. The change in magnetic entropy, $\Delta S_m$, per mole has been calculated from the $M(H)$ curves using Maxwell's thermodynamic relation[53]:

$$\Delta S_m = \int_{H_{initial}}^{H_{final}} \left(\frac{\partial M}{\partial T}\right)_H dH$$

When considering the MCE, it is usual to consider two field regimes: low fields, $\mu_0 H \leq 2$ T, accessible using a permanent magnet and high fields, $\mu_0 H > 5$ T, where the MCE is maximised. Previous work has discussed the role of crystal electric field effects in lanthanide based MCMs with Heisenberg systems performing better at high fields and those with substantial single-ion anisotropy, being suitable for use in lower fields[54,55]. We observe similar behaviour here (Figure S2) and so, consider the MCE of $GdBO_3$ in the limiting field of 9 T and $LnBO_3$, *Ln* = Tb, Dy, Ho, Er, Yb in the low field regime. The change in magnetic entropy, $\Delta S_m$, is typically calculated per unit mole, however for practical applications it is useful to consider per unit mass or per unit volume. The $\Delta S_m$ values at 2 K are given in Table 3 and compared to the standard MCMs, $Gd_3Ga_5O_{12}$ and $Dy_3Ga_5O_{12}$, for fields of 9 T and 2 T respectively[54,56,57]. $GdBO_3$ and $DyBO_3$ are found to have the maximum MCE in gravimetric and volumetric units, in 9 T and 2 T respectively.

Table 3 – Magnetocaloric effect in monoclinic $LnBO_3$ (*Ln* = Gd – Yb) at $T$ = 2 K

| Compound | Field (T) | $\Delta S_m$ (J K$^{-1}$ mol$^{-1}$) | $\Delta S_m$ (J K$^{-1}$ kg$^{-1}$) | $\Delta S_m$ (mJ K$^{-1}$ cc$^{-1}$) |
|---|---|---|---|---|
| **$GdBO_3$** | 9 | 12.5 | 57.8 | 366.3 |
| **$Gd_3Ga_5O_{12}$** | 9 | 14.1 | 41.8 | 296.4 |
| **$TbBO_3$** | 2 | 0.7 | 3.2 | 20.8 |
| **$DyBO_3$** | 2 | 3.1 | 13.9 | 92.5 |
| **$HoBO_3$** | 2 | 0.2 | 0.9 | 6.1 |
| **$ErBO_3$** | 2 | 2.6 | 11.5 | 80.0 |
| **$YbBO_3$** | 2 | 2.1 | 9.1 | 66.0 |
| **$Dy_3Ga_5O_{12}$** | 2 | 3.8 | 11.0 | 80.6 |

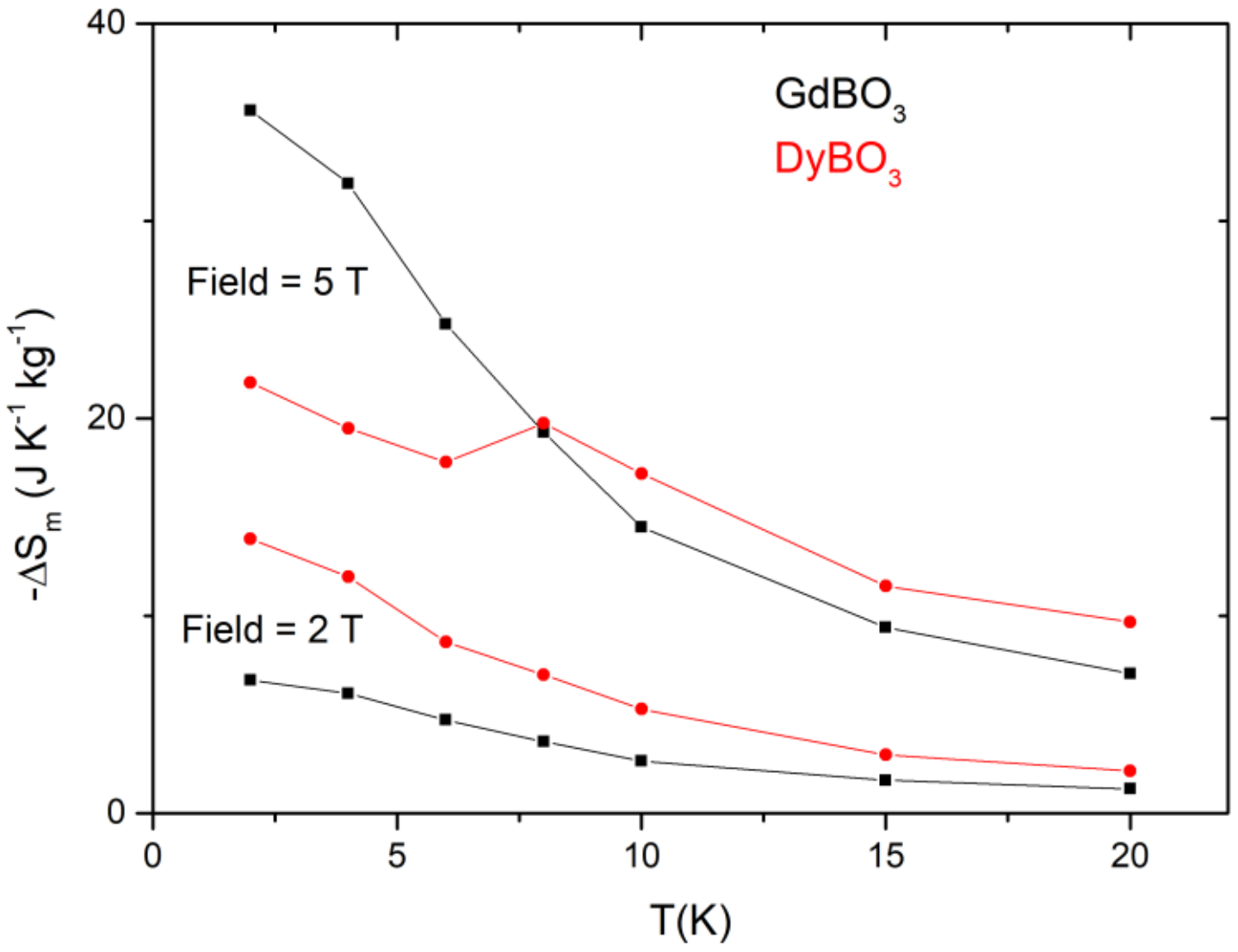

Figure 8 – Change in magnetic entropy $\Delta S_m$ per unit mass as a function of temperature at selected magnetic fields for $GdBO_3$ and $DyBO_3$

How do these materials compare to other low temperature magnetocaloric materials? In recent years, several other $Gd^{3+}$ MCMs have been reported which have high MCE in high magnetic fields[50,58–60]. However, for practical applications, different $Ln^{3+}$ containing compounds can be viable low temperature MCMs in lower fields, up to 2 T[55,57,61,62]. The MCE in $GdBO_3$ and $DyBO_3$ in gravimetric units (Figure 8) are comparable or greater than these materials (Table 3). The origin of the competitive MCE per unit mass can be readily explained by considering the low mass per mole *Ln* ion in $LnBO_3$ (~ 218 g/mol$_{Ln}$) compared to other MCMs $Ln_3Ga_5O_{12}$ (~340 g/mol$_{Ln}$), $LnPO_4$ (~ 255 g/ mol$_{Ln}$), $LnCrO_4$ and $LnVO_4$ (~276 g/ mol$_{Ln}$) and $Ln(HCOO)_3$ (~ 295 g/mol$_{Ln}$). $LnBO_3$ possess the advantage of a low temperature scalable synthesis; this is significant for practical usage. Most importantly, the existence of magnetic ordering transitions for both $GdBO_3$ and $DyBO_3$ at $T < 2$ K means that they would suitable for cooling below 2 K as further magnetic entropy can be extracted below 2 K. Moreover mixed lanthanide orthoborates, $Ln_xLn'_{1-x}BO_3$ could be developed in order to tune the MCE in different temperature and field regimes, as has been reported for other lanthanide systems[55]. $GdBO_3$ and $DyBO_3$ are therefore competitive magnetocaloric materials in the liquid helium temperature regime.

# 4 Conclusion

A series of lanthanide orthoborates $LnBO_3$, *Ln* = Gd, Tb, Dy, Ho, Er and Yb, have been synthesised and their bulk magnetic properties have been measured. They crystallise in a monoclinic structure with the magnetic $Ln^{3+}$ forming a two-dimensional triangular lattice with slight distortions (<1.5%).

Zero field heat capacity measurements reveal different magnetic transitions at $T \leq 2$ K for $LnBO_3$, *Ln* = Gd, Tb, Dy, Er while the onset of magnetic ordering can be seen for $YbBO_3$ at 0.4 K. $HoBO_3$ is postulated to have a non-magnetic ground state. Isothermal magnetisation measurements reveal different single ion anisotropy for the different $Ln^{3+}$. Evaluation of the MCE shows that $DyBO_3$ and $GdBO_3$, are viable magnetocaloric materials in the liquid helium temperature regime in fields $\leq 2$ T achievable using a permanent magnet and higher magnetic fields > 5 T respectively.

The lanthanide orthoborates serve as a prototype of a slightly distorted frustrated rare-earth triangular lattice. We hope that this work will motivate further studies on the nature of the magnetic ground states in these materials.

# 5 Acknowledgements

We acknowledge funding support from the Winton Programme for the Physics of Sustainability. Magnetic measurements were carried out using the Advanced Materials Characterisation Suite, funded by EPSRC Strategic Equipment Grant EP/M000524/1. Supporting data can be found at https://doi.org/10.17863/CAM.13667.

# 7 Supplementary Information

Table S1 – *Ln*-*Ln* bond lengths for monoclinic *Ln*$BO_3$, *Ln* = Gd, Tb, Dy, Ho, Er, Yb

| ***Ln*** | | **Gd** | **Tb** | **Dy** | **Ho** | **Er** | **Yb** |
|---|---|---|---|---|---|---|---|
| ***Ln* –*Ln* (in-plane) (Å)** | 1 | 3.84(2) | 3.789(11) | 3.798(15) | 3.772(8) | 3.782(5) | 3.740(5) |
| | 2 | 3.82(3) | 3.853(15) | 3.78(3) | 3.792(11) | 3.727(6) | 3.720(8) |
| | 3 | 3.846(14) | 3.845(7) | 3.812(10) | 3.808(5) | 3.772(4) | 3.752(4) |
| | 4 | 3.81(2) | 3.746(10) | 3.762(14) | 3.728(7) | 3.753(5) | 3.704(6) |
| | 5 | 3.814(14) | 3.801(7) | 3.772(10) | 3.761(5) | 3.741(4) | 3.713(4) |
| | 6 | 3.88(2) | 3.836(10) | 3.842(14) | 3.821(7) | 3.816(5) | 3.782(6) |
| **<*Ln* – *Ln*> (in-plane)(Å)** | | 3.84 | 3.81 | 3.79 | 3.78 | 3.76 | 3.73 |
| ***Ln*-*Ln* (inter-plane) (Å)** | 1 | 4.32(3) | 4.420(16) | 4.36(3) | 4.365(12) | 4.394(6) | 4.372(9) |
| | 2 | 4.52(2) | 4.449(12) | 4.460(15) | 4.434(9) | 4.405(9) | 4.375(6) |
| **<*Ln* – *Ln*> (inter-plane)(Å)** | | 4.42 | 4.43 | 4.41 | 4.40 | 4.41 | 4.37 |

Figure S1 - Bond angles for the different triangles for the monoclinic *Ln*$BO_3$, *Ln* = Gd, Tb, Dy, Ho, Er, Yb, the deviations for the three sets of triangles are identical. The horizontal dashed line in each plot indicates the angle for ideal equilateral geometry, 60°.

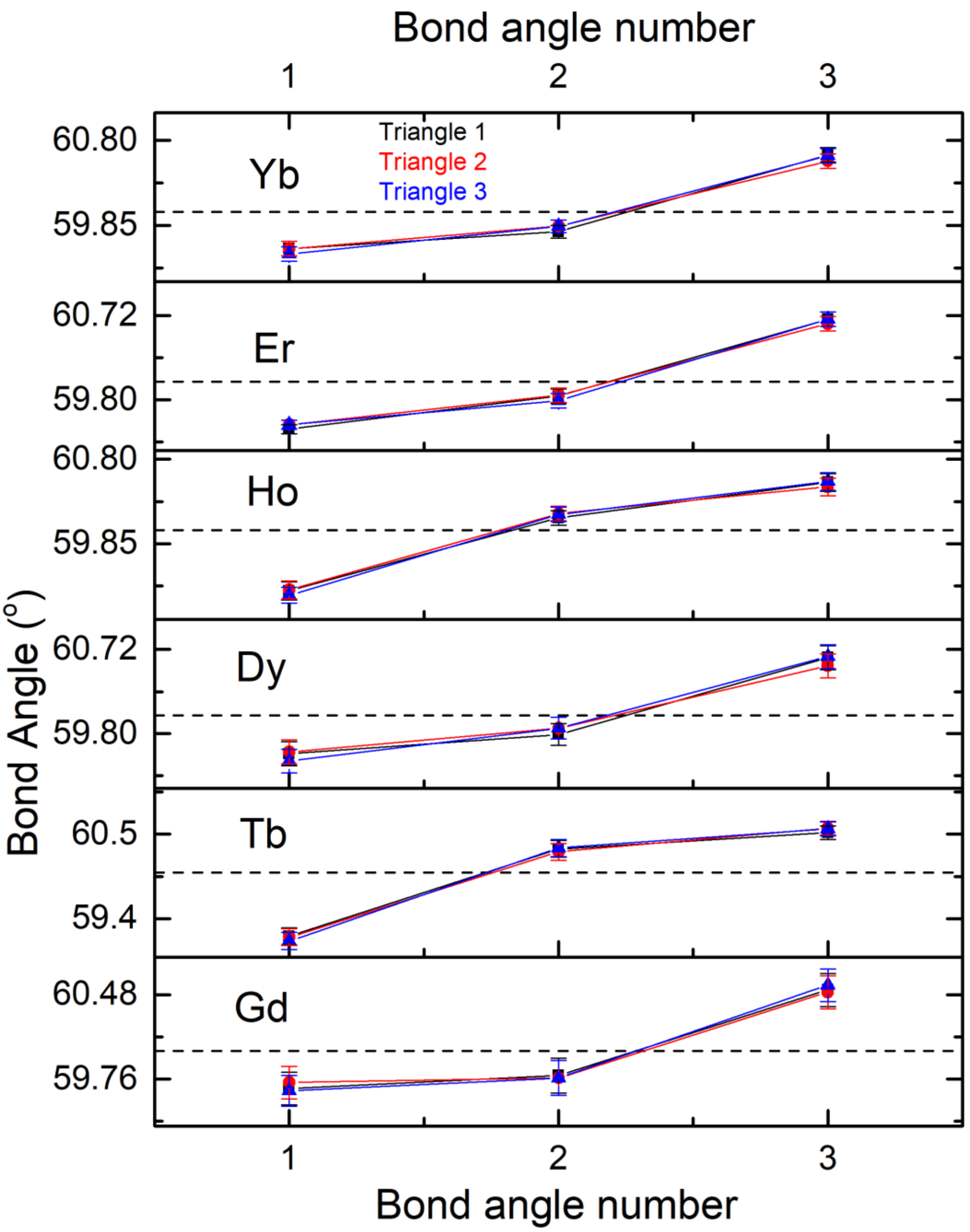

Figure S2 - $\Delta S_m$ (J K$^{-1}$ mol$^{-1}$) for the monoclinic *Ln*BO$_3$, *Ln* = Gd, Tb, Dy, Ho, Er, Yb in the field range $\mu_0 H$ = 0 – 5 T at *T* = 2 K. It is seen that in fields $\mu_0 H \leq 2$ T, attainable by a permanent magnet, DyBO$_3$ has the highest magnetocaloric performance whereas in fields $\mu_0 H > 3.5$ T, GdBO$_3$ surpasses all the other *Ln*BO$_3$ as a magnetocaloric material.

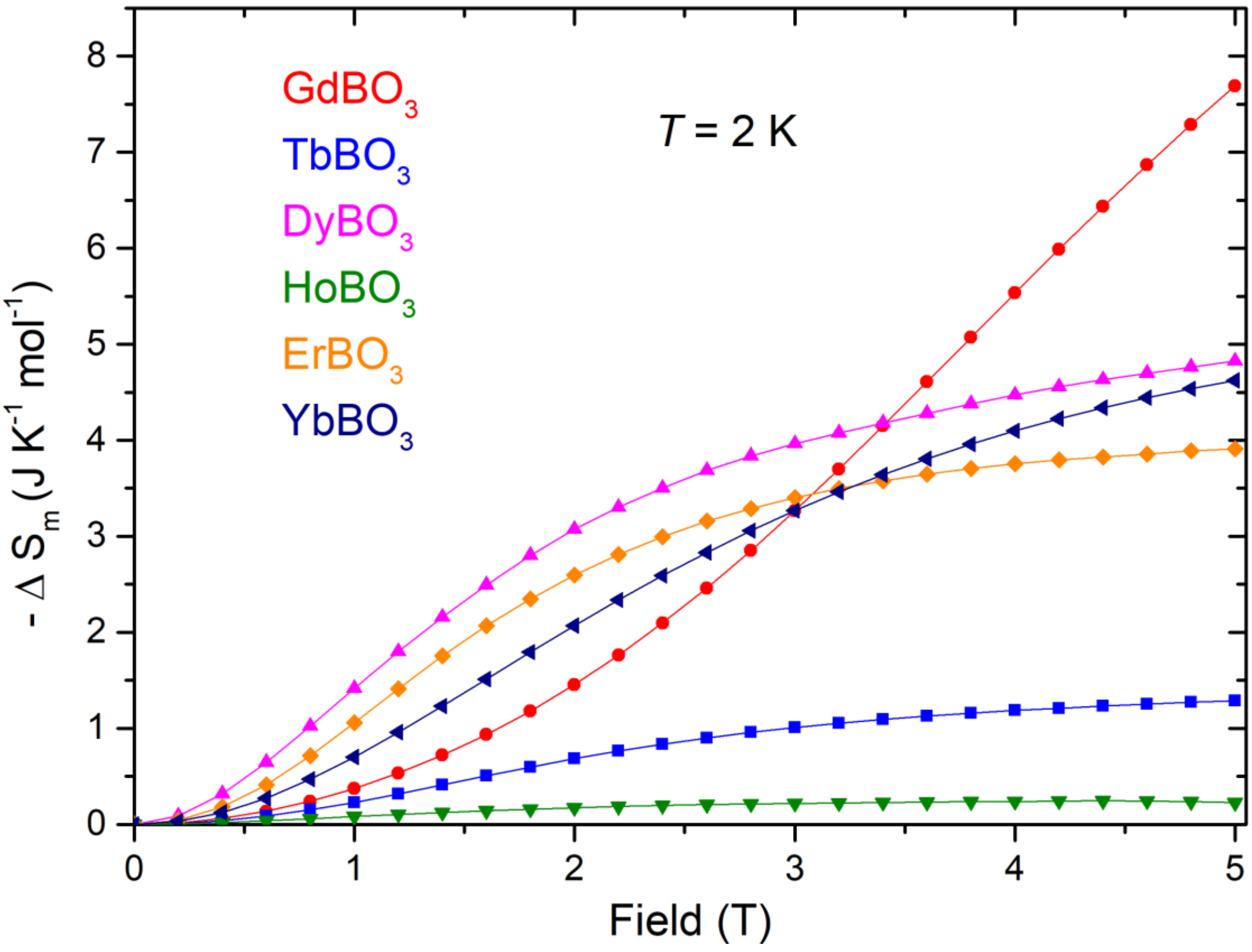